# JUMP DETECTION IN FINANCIAL ASSET PRICES THAT EXHIBIT U-SHAPE VOLATILITY

Cecilia Mancini

Department of Economics, Verona University

21/8/2025

**Abstract**

We describe a Matlab routine that allows us to estimate the jumps in financial asset prices using the Threshold (or Truncation) method of [Man09]. The routine is designed for application to five-minute log-returns, for which microstructure noises are commonly considered negligible. The underlying assumption is that asset prices evolve in time following an Ito semimartingale with, possibly stochastic, volatility and jumps. A log-return is likely to contain a jump if its absolute value is larger than a threshold determined by the maximum increment of the Brownian semimartingale (BS) part. The increments of a BS are particularly sensitive to the magnitude of the volatility coefficient, and from an empirical point of view, volatility levels typically depend on the time of day (TOD), with volatility being highest at the beginning and end of the day, while it is low in the middle.

The first routine presented allows for an estimation of the TOD effect, and is an implementation of the method described in [BolTod2011], appendix B. Subsequently, the routine is used to visualize the TOD effect for the stock Apple Inc. (AAPL).

The second routine presented is an implementation of the threshold method for estimating jumps in AAPL prices. The procedure begins with the construction of a first-round threshold obtained using an initial raw estimate baralpha of the average daily volatility level, and the different TOD factors for the different times of the day. Log-returns are identified to contain a jump when their absolute value exceeds this threshold. The volatility estimate is then updated using only the log-returns not containing jumps, and a second-round threshold is determined. The procedure is repeated recursively until no further jumps are detected. Before application to empirical data, the reliability of the jump detection procedure was tested on simulated asset prices generated by a semi-martingale in which stochastic volatility shows a diurnal effect and leverage, and jumps occur at times generated by a Hawkes process and have Gaussian sizes.

1. **Assumptions and notations.**

p= stock price process; X = log (p); the log-return on the time interval $]t_{j-1}, t_j]$ is defined by ri(j)= Delta_j X = $X_{t_j}$ - $X_{t_{j-1}}$ = log($p_{t_j}$) - log($p_{t_{j-1}}$).

Process X is assumed to evolve in continuous time, following the dynamics

$$dX_t = b_t\, dt + sigma\_t\, dW_t, + dJ_t,$$

where b_t dt is a drift component, the average trend of the asset log price,

sigma_t dW_t is the Brownian component, the multiplier sigma is called **volatility**
dJ_t is the jump component, as described in [Man2009];

the component = b_t dt + sigma_t dW_t is called Brownian semimartingale (BS).
Below, time is measured in financial years: in one year, the stock market is open for 252 days, and on a normal day, the market we consider is open for 6 hours and 25 minutes, which means we have 77 intraday intervals of 5 minutes. Therefore, 5 minutes corresponds to 1/(252*77) financial years.

2. Estimation of the TOD effect using the Bollerslev and Todorov method

**function [TOD,baralpha] = TOD2(m,N,ri,delta)**
% routine that estimates the TOD effect of volatility, sigma, of an asset price.
TOD = time of day effect = factor that must be multiplied by the estimated average daily volatility, to account for its U-shape daily behavior.

% inputs:
**ri**=vector of 5-minute log-returns of the considered financial asset (APPLE in the following application);
**delta**= 5 minutes;
**m**= length(ri) = number of daily returns, without the overnight return (m= 77 in the following application); m represents the number n-1 in [BolTod2011]
**N**= number of days of the data record (N=4815 =length(AAPL)/77 in the following application)
REMARK. The whole observation period [0,T] length is T=N*m*delta, approximately 19 years

% Definition of baralpha = average of BPV_s, where
BPV = BiPower Variation devised in [BarShe2004] for estimating the integrated squared volatility int_0^T sigma^2_u du of a semimartingale with finite activity jumps
BPV_s = sum_{j=2}^{m} |r_j|*|r_{j-1}| is the BPV of day s, which provides an estimate of the average level of squared volatility for day s.
Thus, **baralpha** is the average of the BPV_s across the N days, i.e., an average of the daily squared volatilities

prerim1=ri; % modify the vector of the returns to obtain sum_{s=1}^N BPV_s
prerim1(1)=0;  % change to 0 the first return of every day
for s=2:N
  prerim1((s-1)*m+1)=0;
end

rim1=zeros(1,length(ri)-1);
rim1(1:end)=prerim1(2:end);
riSenzaUltimo=ri(1: length(ri)-1);

```matlab
baralpha=3*sqrt(pi/2)*sqrt( abs(rim1)*abs(riSenzaUltimo) / N );
% abs(rim1)*abs(riSenzaUltimo) = sum_{s=1}^N BPV_s
```

% REMARK. BPV_s approximates sigma^2_{average of day s}*(1 day)
%     (sigma^2 in daily units of measure)*(1 day in daily units of measure)
%     = (sigma^2 in annual units of measure)*(1 day in annual units of measure)
thus, BPV is dimensionless, and the same applies to baralpha

% Construction of NOI_i and TOD_i:
% TOD_i = NOI_i*numerTOD_i/denTOD, the denominator is the same for all i
% NOI_i = numNOI/denNOI(i);  all NOI_i have the same numerator, as i varies

```matlab
Deltan=delta*252; %=1/m
```
% REMARK. [BolTod2011] use Delta_n = 1/(m+1), but to be consistent we put Deltan = 1/m

```matlab
Irit1=( abs(ri)<=baralpha*Deltan^(0.49) );
```
% truncation of the returns with a raw threshold, baralpha is constant. The log-returns below the raw threshold are roughly considered to contain no jumps
```matlab
rit1=ri.*Irit1;
riqt1=rit1.^2;

denTOD= sum(ri.^2);
```
% the denominator of TOD coincides with the Realized Variance (RV) sum_{j=1}^{m*N} r_j^2 on the whole period
```matlab
numNOI=sum(Irit1);
```
%the numerator of NOI counts the number of returns without jumps
```matlab
for i=1:m
  predenNOI=zeros(1,N*m);
  prenumerTOD=zeros(1,N*m);
  for s=1:N
    predenNOI((s-1)*m+i)=Irit1((s-1)*m+i);
```
% indicator showing whether at the fixed time i there was no jump on day s
```matlab
    prenumerTOD((s-1)*m+i)=riqt1((s-1)*m+i);
```
% estimated sigma^2_{t_i} of day s
```matlab
  end
denNOI(i)=sum(predenNOI);
```
% counts the number of days when at time i, no jumps occurred
```matlab
NOI(i)=numNOI/denNOI(i);
```
% proportion of the whole number of returns without jumps over the number of days that at time i had no jumps
```matlab
numerTOD(i)=sum(prenumerTOD);
```
% sum of the squared volatilities at time i across all days. This coincides with denNOI(i) times the average sigma^2_{t_i} across the days where at t_i no jumps occurred
```matlab
TOD(i)=NOI(i)*numerTOD(i)/denTOD
end;
```

**REMARK.** Essentially, TOD(i) is the product of the average sigma^2 at time i, on days when there were no jumps at t_i, with the quotient numNOI/RV.
   If there are no jumps in the sample, numNOI/RV is the reciprocal of the average spot (local)

sigma^2_{t_j} across all days and all times. Thus, at the beginning of the day, usually TOD(i) is bigger than 1, while in the middle of the day, it is less than 1.

If the sample contains some jumps, then the term numNOI/RV has a higher denominator and a smaller numerator. The result is a smaller TOD(i) for every i. The larger the number of jumps occurred, the smaller the factor TOD(i). When a jump occurs, it is likely that most of the price turbulence is explained by the jump itself and that lower volatility is sufficient; thus, it is reasonable for the TOD factor to be lower.

If there exists a time t_i such that every day the i-th return has a jump, then TOD(i) is not defined.

For each time of the day, the TOD factor is the same across days; thus, we have 77 values of TOD(i).

### 3. The TOD effect on Apple's share price

For the stock Apple Inc. we considered 5-minute prices recorded in the Nasdaq U.S. market from 02/01/2003 at 09:30 to 08/03/2022 at 15:55. The overnight returns are excluded, while possible zero returns are retained. We have a number m=77 of 5-minute log-returns per day, for N=4815 days, with a total of 370755 observations. The 5-minute time interval length delta=1/(252*77) is measured in financial years and is constant.

The following chart shows the estimated TOD volatility factor for Apple: when we want to identify a jump in prices, we compare the absolute value of each return with a threshold constructed by multiplying an estimate of average daily volatility by the TOD factor for the time interval over which the return is calculated. In this way, we take into account the characteristic U-shaped pattern of volatility: the threshold is higher at times when volatility is particularly high (typically at the beginning and end of the day), which prevents the incorrect identification of jumps.

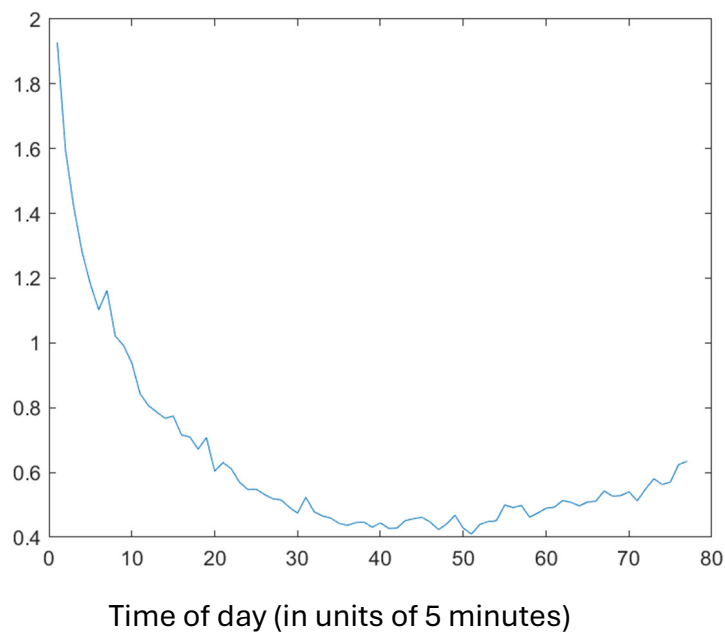

Time of day (in units of 5 minutes)

## 4. Estimating Apple's stock price jumps

A jump in the price p of an asset is detected in the 5-minute interval $]t_{j-1}, t_j]$ if the logarithmic return $ri(j)= \log(p_{t_j}) - \log(p_{t_{j-1}})$ in absolute value is too large compared to a typical increment $b*\delta + \sigma* \Delta_i W$ of the Brownian semimartingale part of the price model.

With probability 1, $\lim_{t_j- t_{j-1} \to 0}$ (maximum increment of W)/mc = 1, where
$$mc^2= 2 * ( t_j- t_{j-1} ) * \log( 1/(t_j- t_{j-1}) )$$
is the squared continuity modulus of the paths of W (see [KarShr1999], p. 114, theorem 9.25 for a precise description). When $t_j- t_{j-1} \to 0$, then
$(\int_{ t_{j-1} }^{t_j} b_u du + \int_{ t_{j-1} }^{t_j} \sigma_u dW_u)^2$ is bounded in absolute value by
$[ \int_{ t_{j-1} }^{t_j} \sigma^2_u du / (_{j-1}}^{t_j}) ]^2 * mc^2$ + higher order terms
(see [Man2009]).

Therefore, the discriminant threshold is set to $\hat{\sigma}_{t_{j-1}}*mc$, where $\hat{\sigma}_{t_{j-1}}$ is an estimate of the spot (i.e., local) volatility coefficient at time $t_{j-1}$. If $|ri(j)|> \hat{\sigma}_{t_{j-1}}*mc$, it is likely that a jump on $]t_{j-1}, t_j]$ occurred.

In turn, the estimation of spot sigma clearly requires that the contribution of jumps be canceled out. The proposed procedure recursively improves the estimation of jumps and local sigma.

**REMARK.** Before applying the jump detection procedure to empirical data, its reliability was tested on simulated asset prices generated by a semi-martingale in which stochastic volatility shows a diurnal effect and leverage, and jumps occur at times generated by a Hawkes process and have Gaussian sizes.

First, we show the function delta_sequence_IndicUni(x,fn0), which is used below to estimate $\sigma_{t_{j-1}}$, given a record of log-returns in a local window adjacent to $t_{j-1}$.

```
function y = delta_sequence_IndicUni(x,fn0)

% This routine allows us to estimate the sigma coefficient locally ([ManMatRen2015]).
To estimate sigma_{t_{j-1}}, it produces a vector of weights f_n(x_ell), where x_ell=t_ell-t((j-1}),
and t_ell belongs to the local window [t((j-1}) , t((j-1}) +h[, where h is a bandwidth.
The delta sequence f_n is generated by a Kernel indicator function, K(x)= I_{0\leq x < 1} and h:
          f_n(x)=K( x/h ) /h = f_n(0)*K( x*f_n(0) ), where f_n(0)=K(0)/h.
The estimator of sigma^2_{t_{j-1}} in AAPLStimaJsConTODfinal.m is defined by
          sum_{ell=1}^m f_n(x_ell))*(ri(ell) truncated)^2,
and corresponds to the average quadratic volatility over a window of length 1 day, ahead

    y = fn0*(x>=0).*(x*fn0<1);
```

## Routine AAPLStimaJsConTODfinal.m

% This routine produces an estimate of the price jumps for the record of 5-minute log-returns stored in AAPL.mat, where zero returns are included, overnight returns are excluded, and the observation interval delta is constant (5 minutes). The estimation is performed with the Threshold Method of [Man09], where the threshold considers the TOD factor estimated by the TOD2(m,N,ri,delta) function above. The parameters m, N, and delta are as in section 1.

```
load('AAPL.mat');
ri=AAPL; %a column vector of 5-minute log returns. We have sqrt(var(ri))=0.0019, mean(ri)=6.8910e-07

m=77; N= 4815;
n=N*m; %n=length(ri)=370755
delta=1/(252*m); % =5.1536e-05
mc= sqrt(2*delta*log(1/delta)); %modulus of continuity of the Brownian motion W paths

[TODAAPL, baralpha]=TOD2(m,N,ri,delta); %calculation of the TOD factor, which must then be multiplied by an estimate of the average daily volatility and by mc, in order to obtain the threshold to be compared with abs(ri)

TODmod=min(1.5, TODAAPL(1:77)); % When the TOD factor is too small, it is replaced with 1.5, as analysis conducted on simulated data has shown that jump detection performance improves with this adjustment. The choice is conservative; it results in an increase in the threshold and, consequently, a decrease in the number of jumps detected.

TODsimulsGlob=zeros(1,n);
for i=1:m
   for s=1:N
      TODsimulsGlob((s-1)*m+i)=TODmod(i);
   end
end % TODsimulsGlob  is a row of m*N components. It contains the list of TOD factors for every observation time t_j. Every day at the same time, the factor is the same.

%% FIRST ROUND

thrPerHatSig=6*baralpha*sqrt(2*delta*log(1/delta)); % Initial raw threshold for truncating log-returns in order to construct the first round estimate of daily sigma
IritPerHatSig=( abs(ri)<=thrPerHatSig );
ritPerHatSig=ri.*IritPerHatSig;
ritqPerHatSig=ritPerHatSig.^2;

% First round estimate of average sigma^2_{t_{j}} for day j. As mentioned above, we calculate a weighted average of 77 returns taken from an adjacent local window, and roughly truncated, with weights given by a delta sequence.
ti=delta*[1:n]; %grid of the times when the prices p_{t_j} are recorded
```

```matlab
fn0= 1/(77*delta);
t=delta*[1:n];
for j=1:N
   hatsigmaq(j) = sum(delta_sequence_IndicUni( ti-t((j-1)*m+1),fn0 ).*(ritqPerHatSig)');
end   % hatsigmaq(j) = average daily sigma for day j

SigmaqGlob=zeros(1,n);
for j=1:N
   for i=1:m
       SigmaqGlob((j-1)*m+i)=hatsigmaq(j);
   end
end % SigmaqGlob is a row of m*N components: at every time t_i of day j, SigmaqGlob assigns the same daily sigma

thr1R=2*TODsimulsGlob.*sqrt(SigmaqGlob)*mc; %first round threshold, this is a row vector; log(1/delta)=9.9115; for day j at time t_i the average daily volatility sigma_j is multiplied by the factor TOD_i
tiJTimes1R= ti'.*(abs(ri)>thr1R'); %column vector containing both non-zero and zero components. A non-zero component i indicates that the corresponding time interval ]delta*(i-1), delta*i] is likely to contain a jump
JTimes1R= delta*find(tiJTimes1R); %"find" returns the position of the non-zero elements in the vector
length(JTimes1R)  %183 = number of detected jumps at the first round

%% SECOND ROUND

ritPer2R=ri.*(abs(ri)<=thr1R'); % the largest jumps are going to be excluded for updating the estimation of sigma
ritqPer2R=ritPer2R.^2;
for j=1:N
   hatsigmaq2R(j) = sum(delta_sequence_IndicUni(ti-t((j-1)*m+1),fn0).*(ritqPer2R)');
end % updated estimation of daily quadratic sigma for every day

Sigmaq2RGlob=zeros(1,n);    % row of m*N components
for j=1:N
   for i=1:m
       Sigmaq2RGlob((j-1)*m+i)=hatsigmaq2R(j);
   end
end

thr2R=2*TODsimulsGlob.*sqrt(Sigmaq2RGlob*2*delta*log(1/delta)); %row of updated threhsolds
tiJTimes2R= ti'.*(abs(ritPer2R)>thr2R');
```

```
JTimes2R= delta*find(tiJTimes2R);
length(JTimes2R), % 3 = number of further detected jumps with the second round
```

%% THIRD ROUND

```
ritPer3R=ri.*(abs(ri)<=thr2R'); % also the 3 returns containing jumps which have been found
with the second round are going to be excluded for updating the estimation of sigma
ritqPer3R=ritPer3R.^2;
for j=1:N
   hatsigmaq3R(j) = sum(delta_sequence_IndicUni(ti-t((j-1)*m+1),fn0).*(ritqPer3R)');
end %row of third round estimates of daily quadratic sigma

Sigmaq3RGlob=zeros(1,n);    % row of m*N components
for j=1:N
   for i=1:m
       Sigmaq3RGlob((j-1)*m+i)=hatsigmaq3R(j);
   end
end

thr3R=2*TODsimulsGlob.*sqrt(Sigmaq3RGlob*2*delta*log(1/delta)); %row of updated thresholds
tiJTimes3R= ti'.*(ritqPer3R>thr3R');
JTimes3R= delta*find(tiJTimes3R);
length(JTimes3R) % 0: no further jumps are found with the third round. The jump detection procedure stops

% We collect all the jumps found
IndSalti=ones(n,1).*(tiJTimes1R>0)+ ones(n,1).*(tiJTimes2R>0);
JTimes1e2R= delta*find(IndSalti);
TotJIdentif= length(JTimes1R)+length(JTimes2R) % 186 identified jump times in total
save 'AAPLhJumpTimesTODfinal', JTimes1e2R
```

%% ANALYSIS OF THE JUMPS FOUND

```
riConJ=AAPL(IndSalti>0); mean(riConJ), sqrt(var(riConJ)),    % -0.0013,  0.0207
hJsizes=riConJ'-sqrt(Sigmaq2RGlob(IndSalti>0))*sqrt(delta); % to estimate the jump size, we subtract the contribution of the Brownian part
mean(hJsizes), sqrt(var(hJsizes)), % -0.0045,  0.0208

hhJsizes=riConJ'-sqrt(Sigmaq2RGlob(IndSalti>0)).*randn(1,length(AAPL(IndSalti>0)))*sqrt(delta);
% estimate the jump size by subtracting the contribution of the Brownian part estimated in a different way
mean(hhJsizes), sqrt(var(hhJsizes)),  % -0.0019,  0.0214
```

```
%% plot
plot(ti, AAPL, '.')
riConJ=AAPL(IndSalti>0);
hold on, plot(JTimes1e2R, riConJ, 'r*')
```

The following picture shows the Apple stock returns (blue points) and the 5-minute intervals estimated to contain a jump (red stars). High returns not detected as jumps indicate moments of particularly high volatility

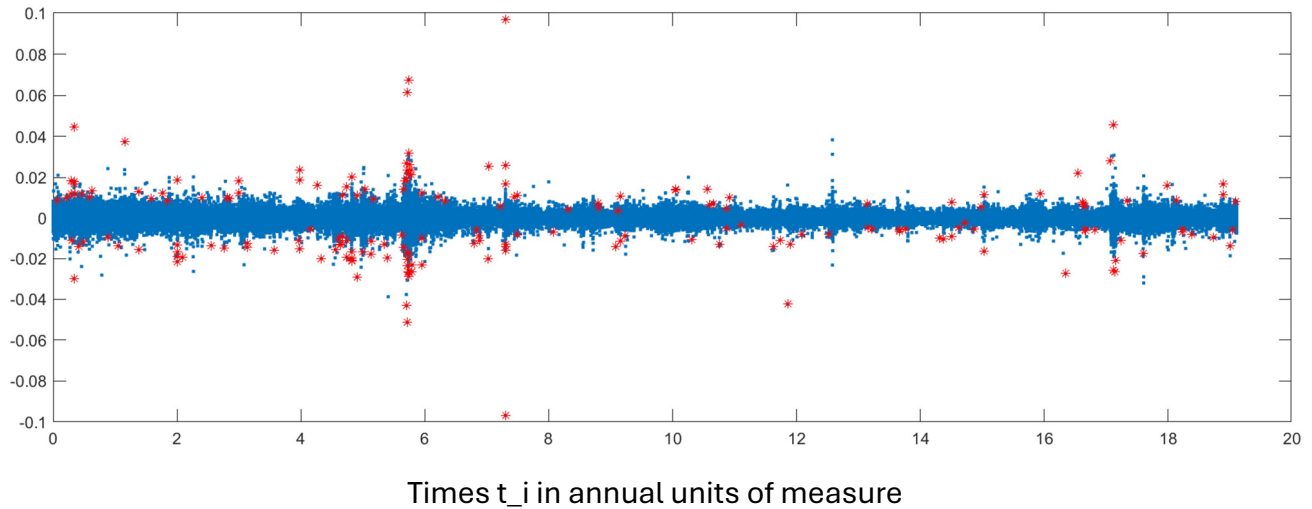

Times t_i in annual units of measure